\documentclass{aa}  
\usepackage{graphicx}
\usepackage{txfonts}
\def\spose#1{\hbox to 0pt{#1\hss}}
\def\simlt{\mathrel{\spose{\lower 3pt\hbox{$\mathchar"218$}}
     \raise 2.0pt\hbox{$\mathchar"13C$}}}
\def\simgt{\mathrel{\spose{\lower 3pt\hbox{$\mathchar"218$}}
     \raise 2.0pt\hbox{$\mathchar"13E$}}}

\def\hi{H\,{\sc i}~}

\def\hii{H\,{\sc ii}~}
\begin{document}

        \title{The effect of clouds in a galactic wind 
               on the evolution of gas-rich dwarf galaxies}
        \author{Simone Recchi\inst{1, 2}\thanks{recchi@oats.inaf.it} 
        \and Gerhard Hensler\inst{1}\thanks{hensler@astro.univie.ac.at}}
        \offprints{S. Recchi}
        \institute{
        Institut f\"ur Astronomie der Universit\"at Wien, 
                 T\'urkenschanzstrasse 17, A-1180 Wien, Austria \and
        INAF - Osservatorio Astronomico di Trieste, Via G.B. Tiepolo 11, 
                34131, Trieste, Italy 
}
        \date{Received  /  Accepted   }
        \abstract
        {Gas-rich dwarf galaxies are probably the closest counterparts
          to primeval objects we can find in the local Universe,
          therefore it is interesting to study their evolution in
          different astrophysical contexts.}
        {We study the effects of interstellar clouds on the dynamical
          and chemical evolution of gas-rich dwarf galaxies. In
          particular, we focus on two model galaxies similar to IZw18
          and NGC1569 in comparison to models in which a smooth
          initial distribution of gas is assumed.}
        {We use a 2-D hydrodynamical code coupled with a series of
          routines able to trace the chemical products of SNeII, SNeIa
          and intermediate-mass stars.  Clouds are simulated by adding
          overdense regions in the computational grid, whose locations
          are chosen randomly and whose density profiles match
          observed ones.  We consider both cloud complexes put at the
          beginning of the simulation and a mechanism for continuous
          cloud formation.  The clouds are inherently dynamically
          coupled to the diffuse gas, and they experience heat
          conduction from a hot surrounding gas.}
        {Due to dynamical processes and thermal evaporation, the
          clouds survive only a few tens of Myr. Due to the additional
          cooling agent, the internal energy of cloudy models is
          typically reduced by 20 -- 40\% compared with models of
          diffuse gas alone.  The clouds delay the development of
          large-scale outflows by mass loading, therefore helping to
          retain a larger amount of gas inside the galaxy.  However,
          especially in models with continuous creation of infalling
          clouds, their bullet effect can pierce the expanding
          supershell and create holes through which the superbubble
          can vent freshly produced metals.  Moreover, assuming a
          pristine chemical composition for the clouds, their
          interaction with the superbubble dilutes the gas, reducing
          the metallicity.  The resulting final metallicity is
          therefore generally lower (by $\sim$ 0.2 -- 0.4 dex) than
          the one attained by diffuse models.  }
   {}

        \keywords{Hydrodynamics -- ISM: abundances -- 
                ISM: jets and outflows -- Galaxies: evolution -- 
                Galaxies: individual: IZw18 -- Galaxies: individual: 
                NGC1569}

\authorrunning{Recchi \& Hensler}

\titlerunning{Effect of clouds on the evolution of dwarf galaxies}

\maketitle

\section{Introduction}
\label{intro}

Gas-rich dwarf galaxies are commonly classified into low
surface-brightness dwarfs, called dwarf irregulars (dIrrs), and higher
surface-brightness objects, usually called blue compact dwarf (BCD)
galaxies.  These classes of galaxies tend to have low metallicities,
blue colors and complex and chaotic gas phases.  A large fraction of
these galaxies shows an ongoing star formation (SF) or at least hints
that this process has been quenched in the recent past.  In this case,
these objects are commonly referred to as {\it starburst} galaxies and
their gas consumption timescales are much shorter than the Hubble time
(Kennicutt \cite{ken98}), making this a transient phase of their
evolution.

Owing to the energy released by stellar winds and supernovae (SNe),
intense episodes of SF are also associated to the development of
galactic winds or at least of large-scale outflows.  The broad
distinction between these two phenomena is the final fate of the
outwards-directed flow of gas: galactic winds generally exceed the
escape velocity while outflows do not, therefore they tend to recede
towards the center of the galaxy.  Clear signatures of outflows are
present in NGC1705 (Hensler et al. \cite{hen98}; Heckman et
al. \cite{hek01}), NGC1569 (Martin, Kobulnicky \& Heckman \cite{mkh02}),
NGC3079 (Cecil et al. \cite{cec01}), IZw18 (Martin \cite{m96}), NGC3628
(Irwin \& Sofue \cite{is96}) among others.  Perhaps the best examples
of large-scale outflows driven by SN feedback are at large redshifts
(Pettini et al. \cite{pet98}; Pettini et al. \cite{pet01}).  Although it
is not certain, in any of the above-mentioned objects, that the metals
will definitely leave the parent galaxy, indirect hints of the
ubiquity of galactic winds are given by the mass-metallicity relation
(Tremonti et al. \cite{tre04}; Dave\'e, Finlator \& Oppenheimer
\cite{dfo06}) and effective yields (Garnett \cite{gar02}).

From a theoretical point of view, the study of the evolution of
gas-rich dwarf galaxies through numerical simulations has been
performed by several authors in the recent past.  The overall picture
is that the occurrence of large-scale outflows is initially driven by
the thermal pressure of a very hot, high pressurized gas and is
favored by a flat distribution of the interstellar medium (ISM), which
allows an easy vertical transport of material.  However, since the
transport of gas along the disk is very limited, outflows are not able
to eject a significant fraction of the ISM, whereas the fraction of
ejected metals can be very large (D'Ercole \& Brighenti \cite{db99};
MacLow \& Ferrara \cite{mf99}; Recchi, Matteucci \& D'Ercole \cite{rmd},
hereafter RMD).  For NGC1569, Martin et al. (\cite{mkh02}) derived a
supersolar metal content in the galactic wind from X-ray spectra but
also advocated mass-loading of it with the ISM.

Most of these studies, however, have focused on flows in homogeneous
media, neglecting the multiphase nature of the ISM, although several
attempts to perform multiphase hydrodynamical simulations have been
made in the past, particularly using the so-called {\it
chemodynamical} approach (Theis, Burkert \& Hensler \cite{tbh92};
Rieschick \& Hensler \cite{rh00}; Hensler, Theis \& Gallagher
\cite{hen04}).  The multiphase nature of the ISM, in particular its
clumpiness, is observationally well established in dwarf galaxies
(Cecil et al. \cite{cec01}; Cannon et al. \cite{cann05}; Leroy et
al. \cite{leroy06}) and it has a solid theoretical background with the
seminal work of McKee \& Ostriker (\cite{mo77}).  According to this
model, the ISM is composed by a cold neutral phase (representing the
cores of molecular clouds), confined by a warm medium (with
temperatures of the order of 10$^4$ K) and these two phases (which are
in pressure equilibrium) are embedded in a hot, diluted intercloud
medium (HIM), continuously produced by SN explosions and stellar
winds.  Sufficiently dense clouds can pierce the HIM without being
swept up, so they can become embedded therein (Vieser \& Hensler
\cite{vh07b}).  At the interface between clouds and HIM,
condensation-evaporation processes establish the final fate of the
cloud and its impact on the development of a galactic wind.

In two previous papers, we have studied the dynamical and chemical
evolution of model galaxies similar to IZw18 (Recchi et
al. \cite{rec04}, hereafter Paper I) and NGC1569 (Recchi et
al. \cite{rec06}, hereafter Paper II).  The main results can be briefly
summarized as follows:

\begin{itemize}

\item most of the analyzed models develop large-scale outflows.  These 
outflows carry out of the galaxy mostly the chemical elements freshly
produced during the most recent episodes of SF, with large escape
fraction of metals with delayed production (like Fe and N).

\item Models with very short burst(s) of SF can cool and mix the newly 
formed metals in a very short timescale, whereas, when the SF is more 
complex, most of the metals are either directly ejected outside the 
galaxy through galactic winds or are confined in a too hot medium, 
therefore cannot contribute to the chemical enrichment of the warm 
ionized medium observed by emission lines from the \hii gas.

\item Models with complex and long-lasting SF episodes reproduce the 
chemical composition and the abundance ratios of the above-mentioned 
galaxies much better than models with bursting SF.

\end{itemize}

In this paper we simulate models with structural parameters similar to
IZw18 and NGC1569.  We increase arbitrarily the gas density of some
specific regions of the computational grid, in order to create a
``cloudy'' phase, and we address the question how and to which extent
a ``cloudy gas phase'' alters the former results.  The clouds possess
a specific density profile and can be either added at the beginning of
the simulation or continuously created during the evolution of the
model.  We then analyze the differences between the dynamical and
chemical evolution of these models with the ones presented in Paper I
and Paper II.  We point out that, at variance with the above-mentioned
works, in this paper we will not specifically look for the best
initial setups and the best assumptions in order to reproduce chemical
and dynamical features of well-known objects.  We will just stress the
main variations produced by a clumpy initial setup.  For this reason,
we will also consider models which failed in Paper I and II at
reproducing the observations of IZw18 and NGC1569.

The paper is organized as follows: in Sect.~\ref{cloud} we briefly
recall the evolution of a cloud embedded in a hot medium; in
Sect.~\ref{model} we present the model and the adopted assumptions in
the simulations.  Results are presented in Sect.~\ref{results_fix}
(models with clouds fixed at the beginning of the simulation) and in
Sect.~\ref{results_inf} (continuous creation of clouds).  Finally, a
discussion and some conclusions are drawn in Sect.~\ref{discussion}.

\section{The dynamics of clouds embedded in a hot phase}
\label{cloud}

The ubiquitous coexistence of cool to warm clouds in the hot phase of
the ISM has attracted during the recent years broad attention on the
interaction of such clouds with this tenuous and hot medium.  The
studies deal with three major effects: the influence of heat
conduction, of shock fronts, and of the dynamics of the flowing hot
gas, like e.g. with or without the presence of large-scale galactic
outflows, on the evolution of clouds (Hartquist et al. \cite{hart86};
Murray et al. \cite{m93}; Ferrara \& Shchekinov \cite{ferr93}; Vietri,
Ferrara \& Miniati \cite{v97}; Fragile et al. \cite{frag04}; Marcolini
et al. \cite{marco05}; Vieser \& Hensler \cite{vh07b} among others).
Despite the very large variety of adopted methodologies, astrophysical
contexts and involved physical processes, and in spite of clearly
defined problems, a broad variety but not yet uniqueness of issues can
be summarized as follows:

\begin{itemize}

\item Moving from the idealized situation treated in analytical 
thermal conduction studies (e.g. Cowie \& McKee \cite{cm77}) to
saturated heat conduction and self-gravitating clouds can change the
results from evaporation to condensation for the same state of the hot
gas and the same cloud mass model (Vieser \& Hensler \cite{vh07a}).

\item A clumpy medium embedded in a hot flow produces mass loading, 
namely the seeding of material, ablated from the clouds, into the
global flow.  It has been demonstrated that this kind of phenomena
helps in clarifying the X-ray emission in starburst galaxies like M82
(Suchkov et al. \cite{suc96}; Strickland \& Stevens \cite{ss00};
Marcolini et al. \cite{marco05}).

\item A single cloud overrun by a shock wave can be crushed within the
so-called crushing time (i.e. the time needed for the internal forward 
shock to cross the cloud and reach its downstream surface) and will be 
destroyed to smaller fragments if cooling dominates. But vice versa it
also evaporates in a parameter regime with exceeding thermal conduction 
(Orlando et al. \cite{orl05}). 

\item In a complex of clouds, if the cloud separation transverse to 
the flow is smaller than some critical value (a few times the typical 
cloud radius, the exact value depending on the authors), the clouds 
will merge into a single structure before the hot flow destroys them.

\item Thermal conduction helps in stabilizing the surface of the cloud, 
making it less susceptible to Kelvin-Helmholtz and Rayleigh-Taylor
instabilities (Orlando et al. \cite{orl05}; Vieser \& Hensler
\cite{vh07b}).  It can also generate an inward-propagating shock wave
able to compress the core of the cloud.

\end{itemize}

In the present work we do not intend to simulate in great detail the
interaction of a cloud or of a cloud complex with a diffuse hot
medium, as made by the previously cited authors.  We instead simulate
galaxy models similar to well observed and studied gas-rich dwarf
galaxies, relaxing the hypothesis of a smooth initial gaseous
distribution (as assumed in Paper I and Paper II) and analyze how the
inclusion of a clumpy medium changes the thermal and chemical
evolution of the ISM.  Indeed, the resolution required to properly
take into consideration conductive fronts surrounding clouds is of the
order of 0.1 pc (Marcolini et al. \cite{marco05}, Vieser \& Hensler
2007a,b), extremely computationally demanding in a simulation in which
the large-scale evolution of the galaxy, up to distances of several
kpc has to be taken into account.

\section{Model description}
\label{model}

\subsection{The numerical code}
\label{model_numerics}

The simulations are performed by means of a 2-D hydrocode in
cylindrical coordinates based on a second-order upwind scheme (Bedogni
\& D'Ercole \cite{bd86}).  The hydro solver is coupled with routines
able to follow in detail the chemical and dynamical feedback on the
galaxy as a consequence of SNeII, SNeIa and winds from
intermediate-mass stars (IMS).

%COMMENT: WHY NOT ALSO HMS?

The chemical evolution has also an impact on the dynamics of the
system, due to the assumption of a metallicity-dependent cooling
function (B\"ohringer \& Hensler \cite{bh89}).  We point out that, as
in the previous papers, when we plot abundances or abundance ratios
produced by our models, we exclude grid points at temperature above 2
$\cdot$ 10$^4$ K.  This is because the gas at these temperatures would
be undetectable with optical spectroscopy and its metallicity could be
guessed only through X-ray analysis, whose use in dwarf galaxies is
still quite uncertain (Martin et al. \cite{mkh02}; Ott, Walter \&
Brinks \cite{owb05}).  The code has been described in detail in RMD and
newer implementations and improvements are reported in Paper I and
Paper II, therefore we refer the readers to these papers for
technicalities.

Given the importance of thermal conduction on the shaping and final
fate of clouds embedded in a hot medium, we just briefly recall the
numerical method adopted to treat this physical phenomenon in our
code.  To solve the heat transport equation, the operator splitting
method is adopted and the one-dimensional problem is solved through
the Crank-Nicholson method (see also D'Ercole \& Brighenti
\cite{db99}).  A saturated heat flux (Cowie \& McKee \cite{cm77}) is
adopted if the mean free path of electrons is larger than the
temperature scaleheight.

\subsection{Model parameters}
\label{model_parameters}

The initial configurations of our models are aimed at reproducing the
main structural parameters of two very well studied gas-rich dwarf
galaxies: IZw18 and NGC1569.  The initial setup is taken from our
previous models and described in RMD (for IZw18) and in Paper II (for
NGC1569).  As described in Sect.~\ref{model_cloud}, the gaseous
distribution is made clumpy, either perturbing the initial setup or
adding clouds as the galaxy evolves, at the same rate as the SF.  We
also vary the initial mass function (IMF) of the stars.  The models
are identified through the notation XYZW, where X refers to the
initial setup (I: setup similar to IZw18; N: setup similar to
NGC1569), Y takes into consideration whether the clouds are put in the
initial setup of the galaxy or are continuously created (B: clouds
present from the beginning; C: continuous creation of clouds).  The
third index refers to the adopted IMF: S is for the Salpeter index
(x=1.35).  A flatter-than-Salpeter index is also tested: A (x=0.95)
and a steeper index (x=1.7) is denoted with K. 

Finally, also the yields from IMS are allowed to change (R for Renzini
\& Voli \cite{rv81}, V for van den Hoek \& Groenewegen \cite{vg97}, and
M for Meynet \& Maeder \cite{mm02}).  For instance, the model called
IBSR starts with a setup aimed at simulating IZw18, puts the clouds at
the beginning of the simulation, assumes the yields of Renzini \& Voli
(\cite{rv81}) for IMS and a Salpeter IMF.  The yields from massive
stars are taken from Woosley \& Weaver (\cite{ww95}), unlike for model
M, where, for self-consistency reasons, the yields of Meynet \& Maeder
(\cite{mm02}) are assumed.  In this case, the upper mass is also 60
M$_\odot$, at variance with the 40 M$_\odot$ value adopted for Woosley
\& Weaver yields.  For the SNeIa, the formulation of Matteucci \&
Recchi (\cite{mr01}) has been adopted, with nucleosynthetic yields
taken from the model W7 of Nomoto, Thielemann \& Yokoi (\cite{nty84}).
The model parameters are summarized in Table~\ref{tab_models}.

As applied in Paper I and Paper II, the SF is gasping (i.e. long
episodes of SF are separated by short periods of quiescence).  More
details about the SF history of various models are given in the
corresponding sections.  We will be referring to {\it diffuse models}
any time we will consider a model, similar to the described cloudy
one, but with a homogeneous initial density distribution, described in
detail either in Paper I or in Paper II.

\begin{table*}[ht]
\caption{Summary of model parameters}
\label{tab_models}
\begin{centering}
\begin{tabular}{ccccc}
  \hline\hline
\noalign{\smallskip}

 Model &  Setup & cloud & IMF slope x & IMS yields$^*$ \\
  &  & creation &  &  \\
\noalign{\smallskip}

  \hline 
   IBSR & IZw18   & no & 1.35 & RV81 \\
   IBSV & IZw18   & no & 1.35 & VG97 \\
   IBAV & IZw18   & no & 0.95 & VG97 \\
   ICSV & IZw18   & yes & 1.35 & VG97 \\
   ICKV & IZw18   & yes & 1.70 & VG97 \\
   NCSM & NGC1569 & yes & 1.35 & MM02 \\
  \hline
 \end{tabular}
\end{centering}

\medskip
$^*$ RV81: Renzini \& Voli (\cite{rv81}); VG97: van den Hoek \& 
Groenewegen (\cite{vg97}); MM02: Meynet \& Maeder (\cite{mm02}).

\end{table*}

\subsection{Cloud description}
\label{model_cloud}

A random generator identifies grid points in the galactic region which
are the cores of the clouds.  In the models in which clouds are put
``ab initio'' in the setup of the model (models identified with the
second index ``B'', see Sect.~\ref{model_parameters}), the central
density is decided a priori (100 cm$^{-3}$), and the total number of
clouds is 25.  The exact number of clouds does not play a significant
role; tests have been performed also with 50 or 75 clouds (reducing
their masses) and this does not affect significantly the results.  The
probability of finding a cloud in a particular grid point is
proportional to the gas density.  Due to the flattened distribution of
the ISM at the beginning of the simulation (see RMD or Paper II), this
gives also a larger probability for the clouds to be found close to
the disk.  The density of the grid points outside the clouds is then
reduced in order to get a consistent final gaseous mass.  In models
with continuous creation of clouds (identified with the second index
``C'') the mass is decided a priori: it is assumed that the clouds are
created at the same rate as the SF rate.

As explained in Paper I, all the stars formed within an interval of
time $\Delta t$ are treated as a single stellar population.  After
this interval of time (typically 10$^6$ yr), a cloud is created, whose
mass equals the mass of the single stellar population.  Moreover,
these clouds are given an initial infall velocity of 10 km s$^{-1}$.
For any of these models, the clouds are approximately shaped according
to a radial density profile $\rho_{\rm cl} \propto R_{\rm cl}^{-1.7}$,
where $R_{\rm cl}$ is the distance from the center of the cloud and
the exponent of the power-law density profile is taken from
observations (de Heij, Braun \& Burton \cite{deh02}; Churchill, Vogt
\& Charlton \cite{chu03}; Tatematsu et al. \cite{tate04}).  It is
important to stress that, given the dimensionality and the symmetry of
our numerical code, the clouds are not spherical, but ring-like
structures.  The clouds are assumed to have primordial chemical
composition (i.e. no metals in them) and the temperature is set to
10$^3$ K.  Since this is also the minimum temperature allowed for the
gas (see Paper I and Paper II), the clouds are not in pressure
equilibrium with the surrounding medium.  In
Sect.\ref{results_IBRS_dyna} we offer arguments demonstrating that
this assumption is not unrealistic and we briefly describe models in
which this assumption is relaxed and clouds are put in pressure
equilibrium with the surrounding medium.

\section{Results of models with fixed clouds.}
\label{results_fix}

\subsection{Model IBSR}
\label{results_IBRS}

\subsubsection{Dynamical evolution}
\label{results_IBRS_dyna}

\begin{figure}[ht]
 \begin{center}
 \includegraphics[width=9cm]{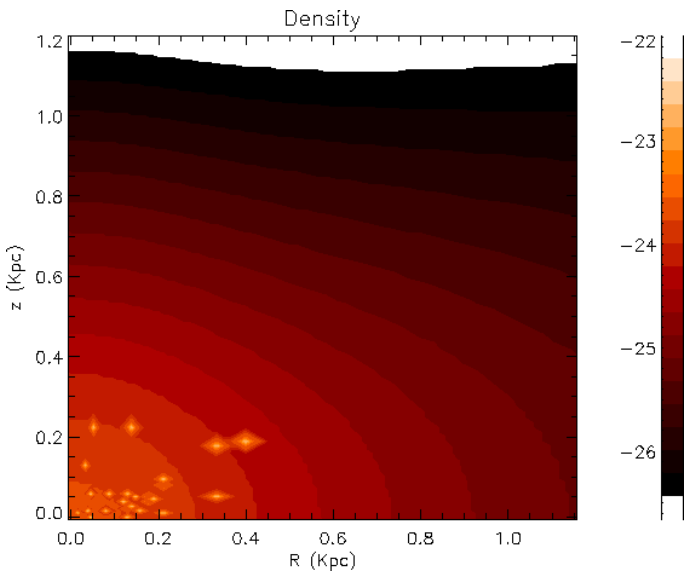}
\caption{Density contours for model IBSR at the 
beginning of the simulation.  The density scale (in g cm$^{-3}$) is 
on the right-hand strip.} 

%COMMENT: SHOULDN'T WE FLIP THE DENSITY SCALING FROM 
%DARK=DENSE TO BRIGHT=DILUTE ? 

\label{setup} % for cross-references 
\end{center}
\end{figure}

\begin{figure}[ht]
 \begin{center}
 \includegraphics[width=9cm]{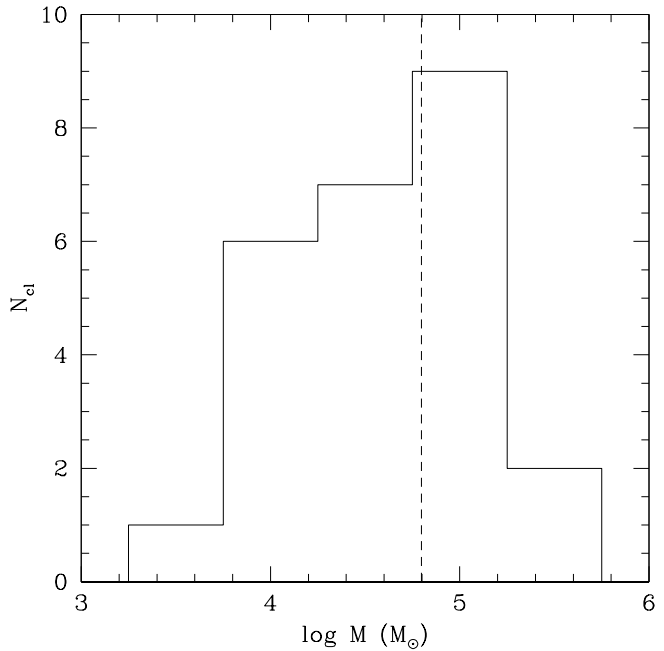}
\caption{Histogram of the logarithmic distribution of the cloud masses 
(in M$_\odot$) for the model IBSR.  The dashed line represents the mean 
value (6.26 $\cdot$ 10$^4$ M$_\odot$).}
\label{masses} % for cross-references 
\end{center}
\end{figure}

We use as a prototypical model with a fixed complex of clouds a setup
similar to the model SR2 analyzed in Paper I.  It has therefore the
same SF history adopted therein, namely a long, moderate episode of SF
lasting 270 Myr, a quiescent period of 10 Myr and a more vigorous
burst (5 times more intense than the first, long-lasting SF episode)
lasting 5 Myr (Aloisi, Tosi \& Greggio \cite{alo99}).  We add randomly
25 clouds, according to the procedure described in
Sect.~\ref{model_cloud}.  The initial gas distribution is shown in
Fig.~\ref{setup}.  It is worth reminding that, since the central
density of the clouds is decided a priori, each cloud has a different
mass, the lightest ones being close to the center.  In
Fig.~\ref{masses} we plot the logarithmic distribution of the cloud
masses.  As we can see, the clouds span almost 2 orders of magnitude
in mass, ranging from $\sim$ 10$^{3.5}$ to $\sim$ 10$^{5.5}$
M$_\odot$, with a mean value of 6.26 $\cdot$ 10$^4$ M$_\odot$.

\begin{figure*}[ht]
 \begin{center}
 \vspace{-3.5cm}
 \includegraphics[width=\textwidth]{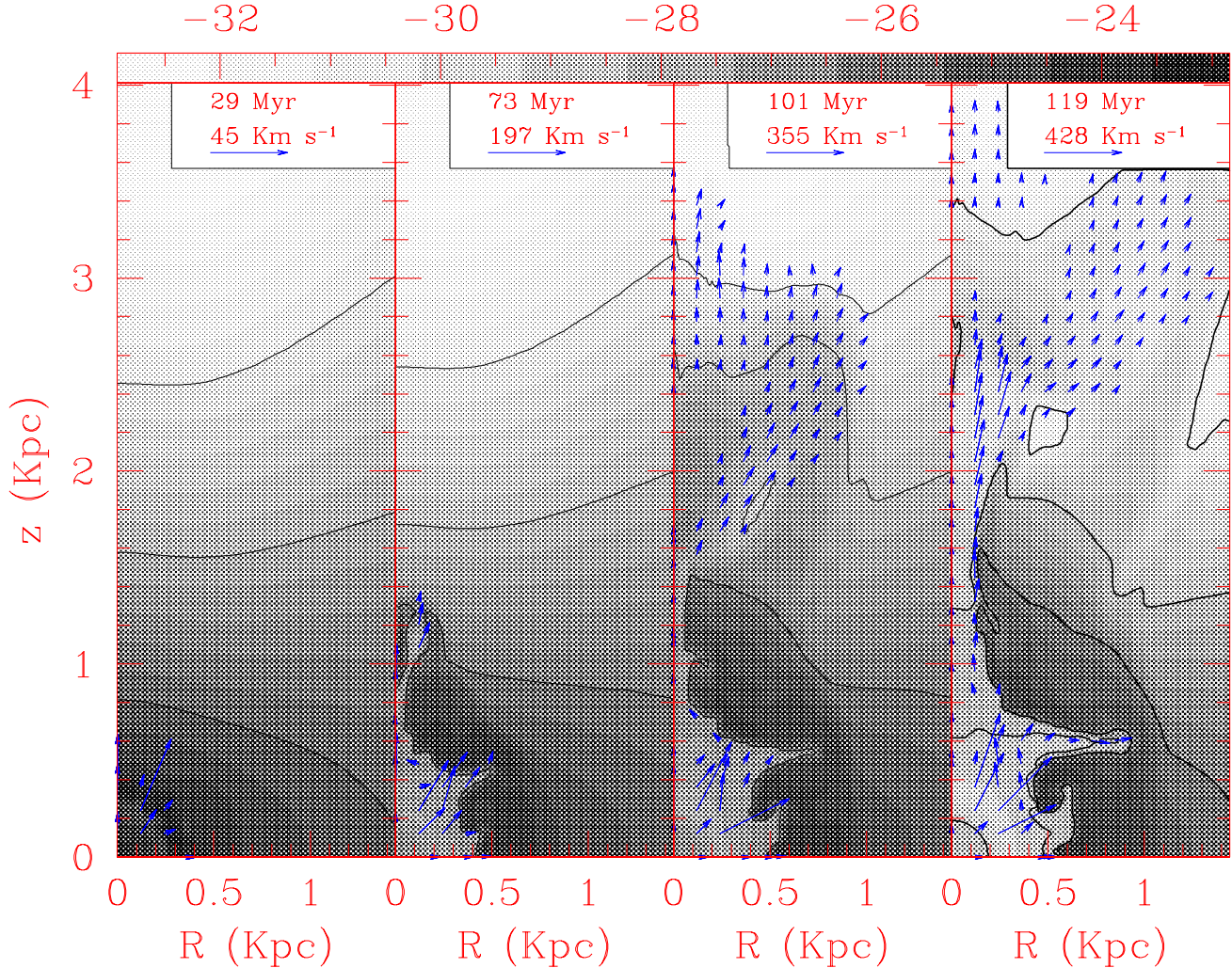}
\caption{Density contours and velocity fields for model IBSR at four 
different epochs (evolutionary times are labeled in the box on top of
each panel).  The logarithmic density scale (in g cm$^{-3}$) is given
in the strip on top of the figure.  In order to avoid confusion,
velocities with values lower than 1/10 of the maximum value (indicated
for each panel in the upper right box) are not drawn.  This is valid
also for Fig.~\ref{t_rho}.}
\label{ibsr} % for cross-references 
\end{center}
\end{figure*}

The evolution of this model in the first $\sim$ 120 Myr is shown in
Fig.~\ref{ibsr}.  This model is able to develop a large-scale outflow
in a timescale of the order of $\sim$ 75 Myr, only slightly delayed
compared to the diffuse model SR2 (for comparison, see figs. 1 and 2 of
Paper I).  This model shows however a much more distorted density
structure and much larger eddies and regions of thermal instabilities.
As we will see better later on, the clouds are destroyed in a
relatively short timescale, but nevertheless they leave an imprint on
the development of the outflow, strongly influenced by the patchiness
of the medium in the central region of the galaxy.

\begin{figure}[ht]
 \begin{center}
 \includegraphics[width=9.5cm]{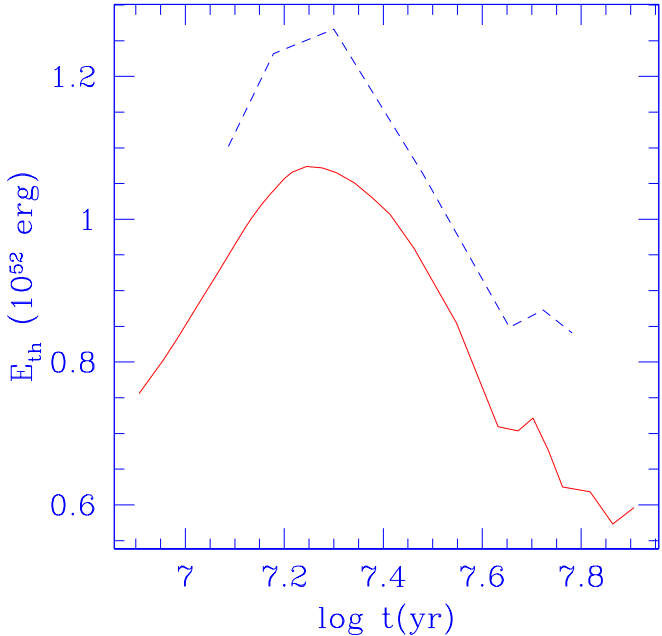}
\caption{Thermal energy (in units of 10$^{52}$ erg) for the model IBSR 
(solid line) and for the reference diffuse model SR2 (Paper I) (dashed
line). }
\label{eth} % for cross-references 
\end{center}
\end{figure}

To quantify this effect, we calculate the total thermal energy of
model IBSR during the first $\sim$ 100 Myr and compare it with the
value found for the model SR2 in Paper I.  The thermal energy is
calculated inside a region $R\leq 1$ kpc and $\leq 730$ pc, which we
have called `galactic region' in RMD and which nearly coincides with
the region where stars are distributed.  This comparison is shown in
Fig.~\ref{eth}.  As we can see, the total thermal energy budget is
clearly affected by the presence of the clouds, leading to a reduction
of $\sim$20\% of the thermal energy deposited into the system, i.e.
it leads to a $\sim$ 20\% increase of the radiative losses.
Incidentally we notice that this value of the total thermal energy
(corresponding to the explosion energy of just a few SNe) is strongly
affected, more than from the radiative cooling of the superbubble,
from our assumption of a low thermalization efficiency of SNeII.  An
extensive discussion about this debated parameter can be found in RMD
and Paper I.

\begin{figure}[ht]
 \begin{center}
 \includegraphics[width=9.5cm]{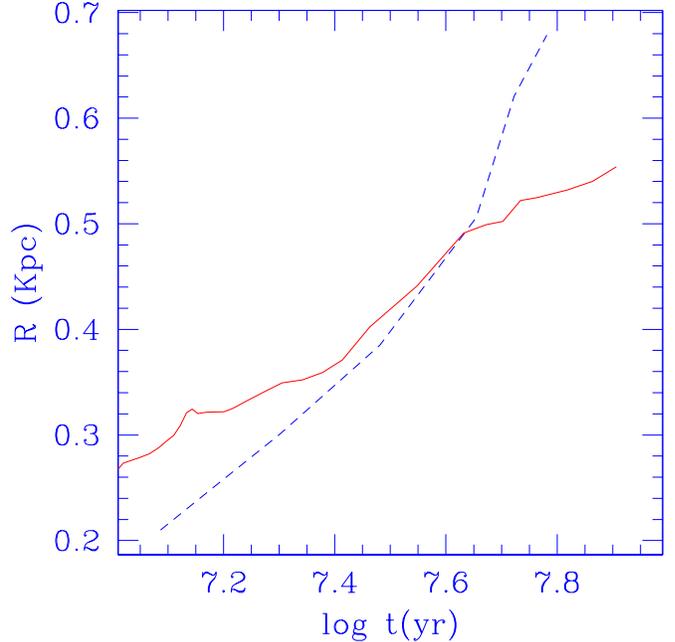}
\caption{Average superbubble radius (in kpc) for the model IBSR 
(solid line) and for the reference diffuse model SR2 (dashed line). }
\label{radii} % for cross-references 
\end{center}
\end{figure}

In Fig.~\ref{radii} we also show the average radius of the superbubble
for the two above-mentioned models.  The cloudy model IBSR allows at
the beginning a slightly faster expansion of the superbubble.  This is
due to the fact that, in order to reproduce the same total mass, in a
cloudy model the density of the diffuse medium has to be reduced.
Moreover, the presence of clouds strongly distorts the shape of the
supershell and the highly pressurized gas inside the cavity can more
easily find regions of lower pressure, from which it is easy to pierce
the shell and break out.  This creates the tongues visible in
Fig.~\ref{ibsr}.  As time goes by, the larger radiative losses
experienced by the model IBSR slow down the expansion of the
superbubble and at $\sim$ 50 Myr the average superbubble radius of the
diffuse model overcomes the one of the cloudy medium.

\begin{figure}[ht]
 \begin{center}
 \includegraphics[width=12.5cm]{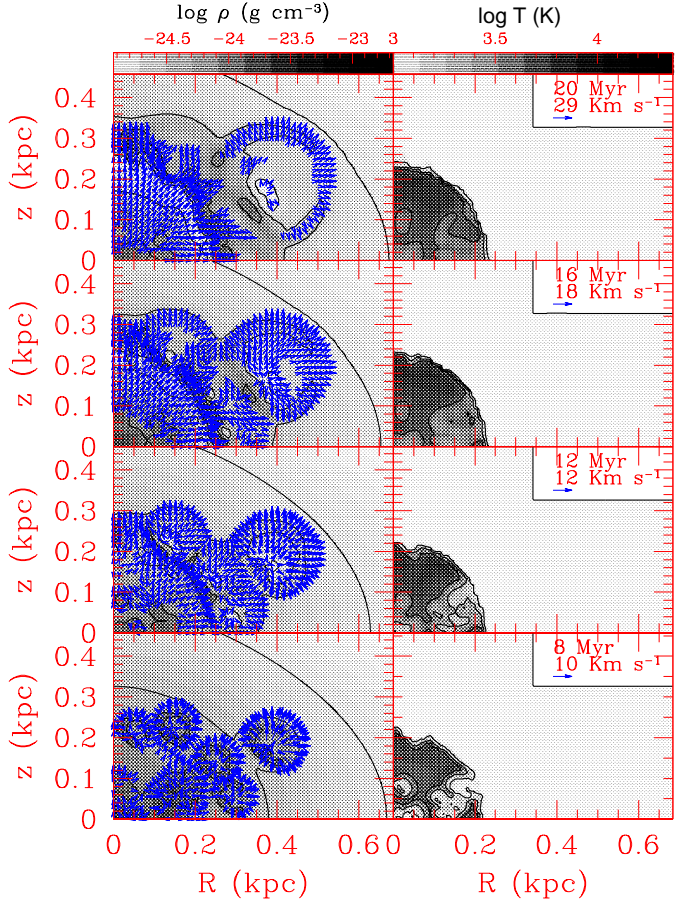}
\caption{ Density contours and velocity fields (left panels) and 
temperature contours (right panels) for the central region of model 
IBSR at four different epochs (evolutionary times are labeled in the 
box on top of each right panel).  Logarithmic scales are given on top 
of each column of panels. }
\label{t_rho} % for cross-references 
\end{center}
\end{figure}

We can better analyze the influence of clouds on the early development
of galactic winds by zooming in the central region of the
computational grid during the first tens of Myr.  A plot of the
density profiles, velocity fields and temperature profiles of the
model IBSR in the first $\sim$ 20 Myr is shown in Fig.~\ref{t_rho}.
We can first notice in this plot that the clouds, even if they are not
encompassed in the superbubble, tend to expand.  This is because they
are not put in pressure equilibrium with the surrounding ISM,
therefore their lifetime is relatively short (a few tens of Myr),
irrespective of the presence of a surrounding HIM.  We have also run
models in which the clouds are put in pressure equilibrium with the
surrounding ISM.  This has been obtained by simply reducing the
temperature of the clouds up to the value which compensate for the
pressure of the ISM at the border of the cloud.  This model shows the
same behavior of the other models once the clouds are encompassed
within the superbubble cavity, namely they are quickly evaporated.
The clouds outside the superbubble show of course a different
behavior, being more stable than the previously described ones, but
this does not affect the global evolution of the model.  Moreover,
growing evidences are accumulating, both observationally
(Ballesteros-Paredes, Hartmann \& V\'azquez-Semadeni \cite{balle99};
Hartmann, Ballesteros-Paredes \& Bergin \cite{hart01}; Hartmann
\cite{hart03}) and theoretically (Elmgreen \cite{elme00}; Heitsch et
al. \cite{hei06}) that molecular clouds are transient structures rather
than well defined objects in quasi-equilibrium states.  Therefore they
must have a relatively short lifetime, as in our simulated clouds.
Given the similarities between the behavior of equilibrium and
non-equilibrium clouds and due to the more complex and slow
computation of equilibrium clouds, we focus from now on only on models
in which the clouds are not in pressure equilibrium.

The clouds create a very patchy temperature distribution in the first
10--12 Myr but the evaporation process continues up to the moment
(after 20 Myr) where the temperature inside the superbubble is almost
uniform.  We can notice once more a supershell which, owing to the
interaction with clouds, strongly deviates from the spherical
geometry, allowing therefore larger shears and eddies (and
consequently a reduced thermal energy of the bubble), but also tongues
and fingers from which it is easier to leak out of the superbubble.
This is the reason why, after $\sim$ 100 Myr some outflowing gas is
already $\sim$ 3 kpc above the galactic disk (third panel of
Fig.~\ref{ibsr}) although the average superbubble radius is at this
stage smaller than for the diffuse model SR2 (Fig.~\ref{radii}).

\subsubsection{Chemical evolution}
\label{results_IBRS_chem}

\begin{figure}[ht]
 \includegraphics[width=9.5cm]{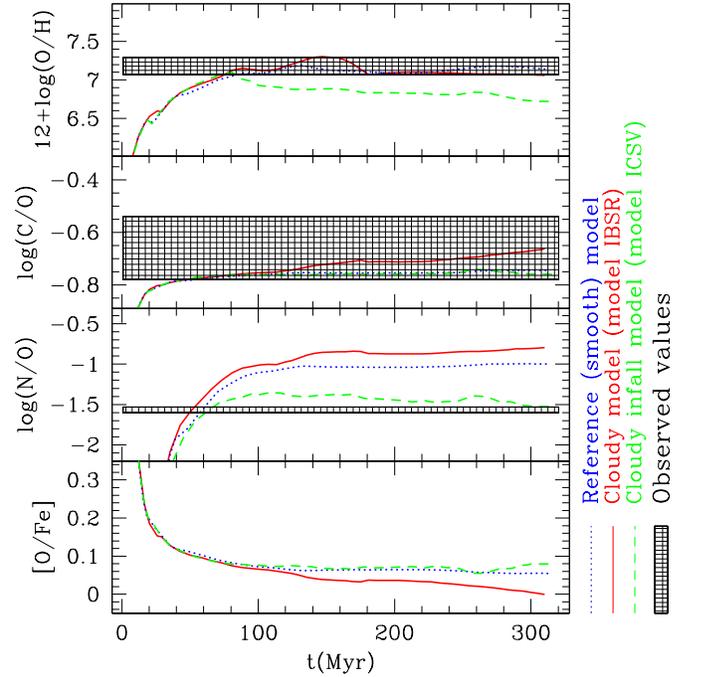}
\caption{ Evolution of 12 + log (O/H) (top panel), log (C/O) (second 
  panel), log (N/O) (third panel) and [O/Fe] (bottom panel) for a
  prototypical model with a fixed initial cloud complex (IBSR model)
  (solid line) and a prototypical model with continuous creation of
  cloud (ICSV model) (dashed line).  These models are compared with a
  model of similar mass but smooth ISM distribution (dotted line).
  The superimposed shaded areas indicate the observed values found in
  literature (if available), with relative error-bars.}
\label{cno1}
\end{figure}

From a chemical point of view, the expected effect of clouds is to
dilute the hot metal-rich gas through evaporation of the (metal-poor)
clouds, allowing for a reduction of the metallicity without altering
the abundance ratios (K\"oppen \& Hensler \cite{kh05}).  However, the
reduced thermal energy and, consequently, the reduced escape fraction
of metals from the galactic region (Sect.~\ref{results_IBRS_dyna})
should produce an {\it increase} of the metallicity of the galactic
ISM.  In this case, the abundance ratios are affected if the ejection
efficiencies depend on the different chemical species (here we simply
define ejection efficiency as the fraction of metals outside the
galactic region compared to the total amount which has been
synthesized).  To disentangle these two competing effects, we analyze
the differences in the chemical evolution between diffuse and cloudy
models.

The comparison of 12+log(O/H), log(C/O), log(N/O) and [O/Fe] of the
warm ionized phase is presented in Fig.~\ref{cno1}.  In this plot we
also show the evolution of a prototypical model of continuous creation
of cloud (model ICSV, see Sect.~\ref{results_ICSV_chem}).  The
evolution of $\alpha$-elements is not significantly altered by the
presence of clouds (the difference being always around $\sim$ 0.1
dex), but we can notice less negligible differences (of the order of
0.2 dex) in the log(N/O) abundance ratio.  Indeed the diffuse model,
due to the reduced radiative losses, attains a larger fraction of
metals with temperature above 2 $\cdot$ 10$^4$ K, therefore excluded
by this plot.  Nitrogen is mostly produced during the thermal pulsing
phase by AGB stars of masses $\sim$ 4 -- 7 M$_\odot$ therefore with a
delay compared to the prompt production of oxygen.  Soon after its
production, nitrogen is also located inside a hot cavity (carved by
SNe) most likely than oxygen.  Moreover, as demonstrated in RMD, the
ejection efficiency of nitrogen can be larger than the one of
$\alpha$-elements, again favoring the decrease of N/O in the diffuse
model, for which the development of a large-scale outflow is
anticipated.

%COMMENT: I DO NOT UNDERSTAND ''EJECTION EFFICIENCY''. DO YOU MEAN 
%FRACTION EJECTED?

Incidentally, we can notice that the assumption of a cloudy medium
worsens the agreement between the predicted log (N/O) and the
observations.  However, as we have pointed out in Paper I, the Renzini
\& Voli (\cite{rv81}) yields tend to overestimate the nitrogen
production and only the assumption of the Meynet \& Maeder
(\cite{mm02}) set of yields can reconcile the prediction of the models
with the observed abundance ratios in IZw18.  We stress once again
that the main goal of this paper is a study of the effect of a cloud
complex on the dynamical and chemical evolution of a gas-rich dwarf
galaxy rather than the attempt to exactly reproduce the chemical
features of specific objects.

\subsection{Model IBSV}
\label{results_IBRV}

\subsubsection{Dynamical results}
\label{results_IBSV_dyna}

\begin{figure*}[ht]
 \begin{center}
 \hspace{-2.6cm}\includegraphics[width=16cm]{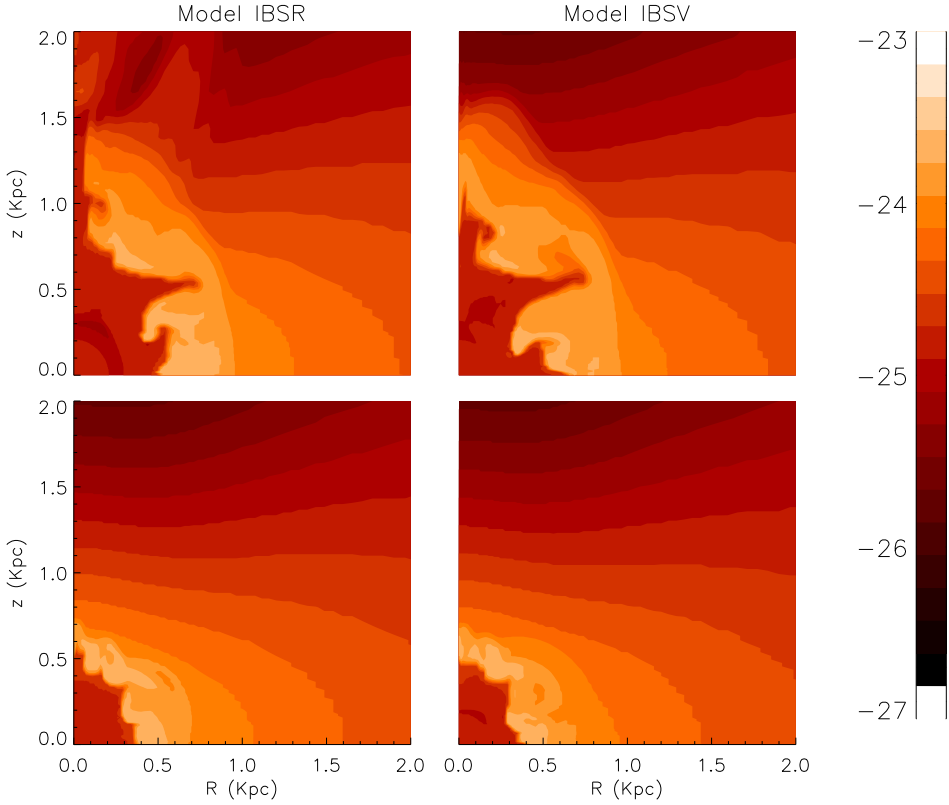}
\caption{ Density contours for model IBSR (left panels) and model 
IBSV (right panels) at two evolutionary times: 50 Myr (lower panels) and 
100 Myr (upper panels).  The density scale (in g cm$^{-3}$) is 
on the right-hand strip.  }
\label{snap_comp} % for cross-references 
\end{center}
\end{figure*}

The model IBSV differs from the model IBSR only on the assumed IMS
yields (van den Hoek \& Groenewegen \cite{vg97} instead of Renzini \& Voli
\cite{rv81}; see Table~\ref{tab_models}).  Dynamically, the only difference
is therefore a variation of the cooling rates, due to our assumed
metallicity-dependent cooling function.  It is interesting to quantify
this effect through a direct comparison of models differing solely in
the adopted nucleosynthetic yields.  This comparison is shown in
Fig.~\ref{snap_comp} at two evolutionary times: 50 Myr (lower panels)
and 100 Myr (upper panels).  Overall, the agreement between the
dynamics of these two models is very good.  However, the extremely
non-linear character of superbubble evolution can be noticed in this
plot and small differences in the energy budget of the model (due to
different cooling rates) can produce significant changes in the
dynamics.  In particular, in model IBSR after 100 Myr some gas is
already flowing out of the galaxy through a narrow nozzle, whereas in
model IBSV the occurrence of a large-scale outflow is slightly
delayed.  At this point, the total mass of gas within the galaxy is
$\sim$ 9\% larger in model IBSV, a small but non-negligible factor.

\subsubsection{Chemical results}
\label{results_IBSV_chem}

\begin{figure}[ht]
 \includegraphics[width=9.5cm]{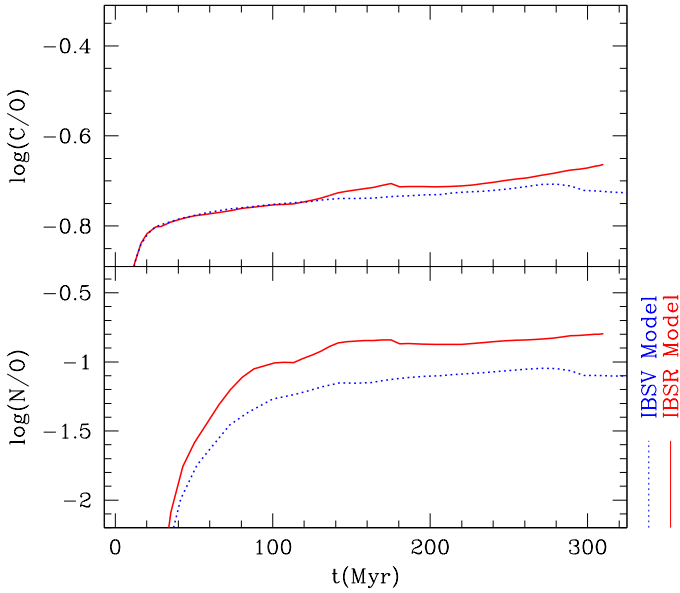}
\caption{ Evolution of log (C/O) (upper panel) and log (N/O) (lower 
        panel) for models IBSR (solid line) and IBSV (dotted
        line).}
\label{cn_rv_vg}
\end{figure}

The comparison of log(C/O) and log(N/O) in models IBSR and IBSV is
shown in Fig.~\ref{cn_rv_vg}.  In this plot we do not show the
evolution of oxygen because its production in IMS is negligible.
Since the prescriptions for the yields from massive stars are the same
(Woosley \& Weaver \cite{ww95}), we do not see significant differences
in the two models.  The yields of van den Hoek \& Groenewegen
(\cite{vg97}) produce substantially less carbon and nitrogen compared
to Renzini \& Voli (\cite{rv81}) yields (Chiappini, Romano \& Matteucci
\cite{crm03}).  The difference is particularly significant for what
concerns N, whose production is approximately halved, resulting
therefore in a N/O $\sim$ 0.3 dex lower than in the model IBSR.  The
same difference has been produced by diffuse models with different IMS
yields (e.g. Recchi et al. \cite{rec02}; Paper I), indicating that the
different dynamics of the cloudy model do not significantly affect
this behavior.

\subsection{Varying IMF: model IBAV}
\index{results_flatter}

As shown in Table~\ref{tab_models}, a model similar to IBSV but with
flatter IMF is also considered.  This models has the same SF history
considered so far (i.e. the one derived from the work of Aloisi et al.
\cite{alo99}), but the energy injection rate is much larger.  To be
more precise, the total energy release is $\sim$ 2.5 times larger than
in IBSV.  Since the gas binding energy remains the same, in spite of
the larger radiative losses due to the interactions clouds-HIM, this
energy release is enough to unbind all the gas initially present in
the galactic region (an ellipsoid of dimensions $\sim$ 1000 $\times$
730 pc, see RMD) in $\sim$ 250 Myr.  This complete blow-away does not
happen in IBSV, where a large-scale outflow occurs in the polar
direction but most of the gas close to the disk remains in the
galactic region.

\section{Results of models with continuously created clouds.}
\label{results_inf}

As described in Sect.~\ref{model_cloud}, in this set of models we
produce a cloud each $\Delta t$ yr (typical value 10$^6$ yr), having a
mass equal to the total amount of gas turned into stars in the same
interval of time.  In the framework of simple closed-box models of
chemical evolution, this case, (infall rate equal to the SF rate) is
called {\it extreme infall} (Larson \cite{lar72}) and leads to the
simple expression for the metallicity $Z = y_Z [1 - e^{-(\mu^{-1} -
1)}]$, where $\mu$ is the gas mass fraction and $y_Z$ is the total
yield (i.e. the ratio between the total mass in metals newly formed
and the amount of mass locked up in low mass stars and remnants).  The
relaxation of the instantaneous recycling approximation and the
inclusion of dynamical effects (winds and mixing and cooling of
metals) changes this finding but K\"oppen \& Edmunds (\cite{ke99})
demonstrated that the ratio between infall and SF rate is the
determining factor in the chemical evolution of galaxies.  The clouds
are given an infall velocity of 10 km s$^{-1}$ along the polar
direction, their location in the computational grid is randomly chosen
and their profile is again $\rho_{\rm cl} \propto R_{\rm cl}^{-1.7}$,
but in this case the central density is constrained by the total mass
of the cloud and by its location rather than being constant for each
cloud.

\subsection{Model ICSV}
\label{results_ICSV}

\subsubsection{Dynamical results}
\label{results_ICSV_dyna}

\begin{figure*}[ht]
 \begin{center}
 \includegraphics[width=\textwidth]{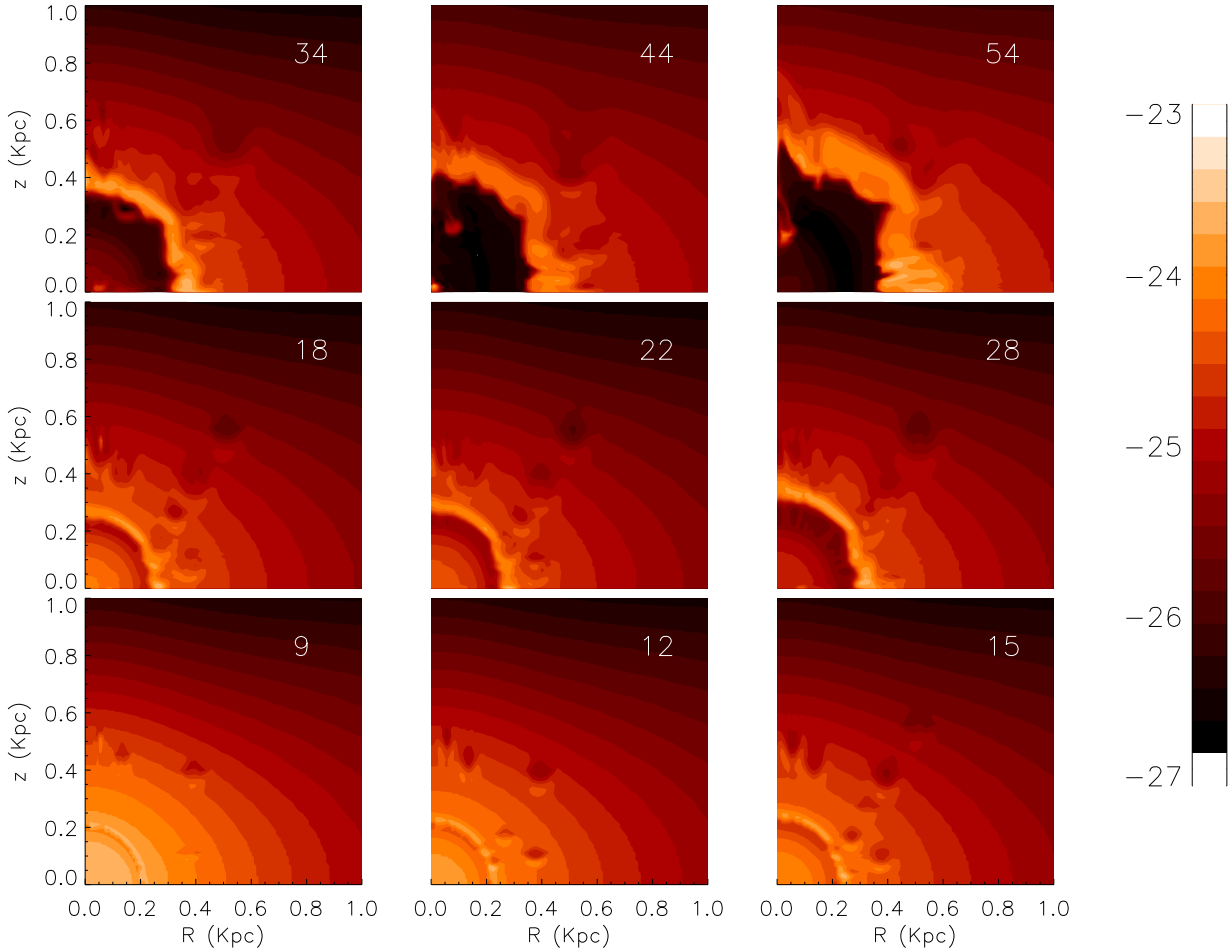}
\caption{ Density contours for the warm gas for model ICSV at 9 evolutionary 
times (labeled in Myr at the top right corner of each panel). The
density scale (in g cm$^{-3}$) is on the right-hand strip.  }
\label{9s} % for cross-references 
\end{center}
\end{figure*}

As prototype of this group of models, we use a setup similar to IBSV,
the only difference being the mechanism of cloud formation.  Given
the assumed SF history, during the first episode the clouds have a
mass of 6 $\cdot$ 10$^3$ M$_\odot$, which increases to 3 $\cdot$
10$^4$ M$_\odot$ during the last burst.  Snapshots of the evolution of
this model in the first $\sim$ 55 Myr are shown in Fig.~\ref{9s}.  In
this figure the shocks created by the clouds in their descent to the
galactic disk are quite evident, in particular in the bottom row of
panels.  In particular, a bow shock is created around the cloud and a
reverse shock is generated downstream behind the cloud, leaving an
underdensity region behind it.  The structure is also highly
Kelvin-Helmholtz unstable.  The timescale for the growth of
Kelvin-Helmholtz instabilities is approximately

\begin{equation}
t_{\rm K-H} = {q^{0.5} \over {k v_{\rm inf}}},
\end{equation}
\noindent
(Chandrasekhar \cite{cha61}) where $q$ is the ratio between the cloud
and the intercloud densities, $v_{\rm inf}$ is the infall velocity of
the cloud (relative to the local ISM) and $k$ is the wavenumber of the
unstable mode.  The most unstable models are the ones with $k \sim
r_{\rm c}^{-1}$ (where $r_{\rm c}$ is the radius of the cloud),
leading to a $t_{\rm K-H}$ between 10 and 20 Myr with our parameters
(depending on the size of the clouds, which is not constant).  This is
therefore also the timescale for the fragmentation of the cloud and
its mixing with the local ISM.  At later times, a non-negligible
probability exists, given our simplified assumptions, that a cloud is
created directly inside the expanding superbubble.  An example is
visible in the upper right panel of Fig.~\ref{9s} (at (R, z) $\sim$
(20, 200) pc).  In this case, given the much larger density ratio $q$
(between 10$^4$ and 10$^5$), the Kelvin-Helmholtz timescale becomes
larger than the time considered in our simulations (a few hundreds of
Myr).  The cloud is therefore ablated by the flow of gas pushed by the
exploding SNe (creating mass loading) and evaporated by the large
temperature of the cavity (few 10$^6$ K) in a timescale of the order
of few tens of Myr.  Occasionally, clouds are created close enough in
space and time, such that mutual interaction between clouds manifests.
In this case, the clouds form coherent structures like the ones
described by Poludnenko, Frank \& Blackman (\cite{polu02}) before
being evaporated.

Due to the combined effect of thermal evaporation and bow shocks, the
thermal energy budget is still smaller (by $\sim$ 15\%) than for
model IBSR (plotted in Fig.~\ref{eth}).  Owing to the larger radiative
losses and to the ram pressure of the infalling clouds, a galactic
wind develops after $\sim$ 130 Myr, delayed compared to model IBSR.
However, as pointed out in Sect.~\ref{results_IBRS_dyna}, the infall
of clouds can help in structuring and fingering the supershell,
creating ways out for the hot gas.  Indeed, the ram pressure of the
infalling clouds is $p_{\rm ram} = \rho_{\rm ISM} \cdot v_{\rm inf}^2$
$\sim$ 10$^{-11}$ erg cm$^{-3}$, of the order of the ram pressure of
the expanding supershell and larger than the thermal pressure of the
hot cavity.  Moreover, it is worth pointing out that, for this set of
models, the total gaseous mass is kept constant by the assumption that
the cloud creation rate should balance the SF rate (see
Sect.~\ref{model_cloud}), at variance with model IBSR, in which a
fraction of gas (at a rate of 6 $\cdot$ 10$^{-3}$ M$_\odot$ yr$^{-1}$,
see Paper I) is continuously turned into stars.

\subsubsection{Chemical results}
\label{results_ICSV_chem}

In Paper II (sect. 4.7) we have analyzed the effect of the infall
(along the polar direction) of a very large and very massive cloud.
In this case, the development of an outflow is completely hampered and
all the metals freshly produced by the ongoing SF are trapped inside
the galactic region, preventing the development of differential
outflows, a natural outcome of this kind of simulations.

As we have seen, in model ICSV the ``cap'' effect is less significant:
it helps reducing the total thermal energy able to drive a large-scale
outflow, but the clouds are dissolved on a timescale of the order of a
few tens of Myr.  Moreover, as seen in Sect.~\ref{results_ICSV_dyna},
they can pierce the supershell, creating funnels for the free flow of
the hot, high-pressure gas.  They can therefore (slightly) delay the
development of an outflow, but they cannot prevent it.  Consequently,
the chemical evolution of this model is strongly affected by the
dilution effect of the clouds.  In particular, the process of cloud
dissolution described in Sect.~\ref{results_ICSV_dyna} continuously
allows a mixing of the metals with pristine gas.  This effect results
to be, in our simulation, much larger than the ``cap'' effect of the
infalling clouds.  We can see the chemical evolution of model ICSV in
Fig.~\ref{cno1}.  The final oxygen abundance is $\sim$ 0.3 dex smaller
that the oxygen attained by model IBSR.  Even more important is the
fact that, after $\sim$ 90 Myr, the oxygen abundance mildly but
constantly decreases as function of time.  This is due to the fact
that the selective loss of metals has not been suppressed and that the
continuous creation (and subsequent disruption) of clouds mixes the
ISM with unpolluted gas.  Very significant is also the effect on the
N/O abundance ratio (a difference of $\sim$ 0.6 -- 0.7 dex), but we
stress once again that in this case the main reason of this difference
is the choice of IMS yields.  As we can see from Fig.~\ref{cn_rv_vg},
the final log(N/O) of model IBSV is $\sim$ 0.3 -- 0.4 dex larger than
model ICSV.

We also tested a model with a larger initial total mass of gas at the
beginning of the simulation.  The setup of this model is similar of
model SV3 of Paper I (i.e. a total initial mass of 3 $\times$ 10$^7$
M$_\odot$ instead of the standard value of 1.7 $\times$ 10$^7$
M$_\odot$), but in it we apply the same procedure of continuous
creation of clouds analyzed for model ICSV.  The slightly lower
density contrast parameter $q$ does not significantly affect the
overall cloud-ISM interaction process and the dissolution timescale of
the clumps but, due to the larger ISM pressure, the development of a
large-scale outflow is largely delayed.  It occurs only $\sim$ 250 Myr
after the beginning of the SF process.

\subsection{Model ICKV}
\label{results_ICKV}

\subsubsection{Dynamical results}
\label{results_ICKV_dyna}

This model has the same setup and same SF history of model ICSV, the
only difference being a steeper (x=1.7) IMF.  Consequently, the energy
return rate is much smaller than the above-considered model (about a
factor $\sim$ 2) and, since the binding energy of the gas is the same,
the reduced power of the burst has deep consequences on the
development of a large-scale outflow.  In this model, a break-out of
the superbubble with consequent outflow of gas happens at around
$\sim$ 180 Myr, but its intensity is very mild and the continuous
infall of clouds is sufficient to suppress its further development and
to close the funnel.  The final structure (after $\sim$ 300 Myr) is an
elongated ellipsoid of $\sim$ 700 $\times$ 220 pc, with just some
traces of gas which has managed to leak out and flows freely, mainly
along the polar direction. Later on, no SF is occurring anymore,
therefore the process of cloud formation is suppressed.  Although the
input of energy is continuous in these models (due to the contribution
of SNeIa), the produced energy is not enough to break-out again,
therefore the supershell tends to recede towards the center of the
galaxy (see Recchi \& Hensler \cite{rh06}).  Only $\sim$ 3\% of the
gas produced during the two episodes of SF leave the galactic region
at the end of the simulation.

\subsubsection{Chemical results}
\label{results_ICKV_chem}

The chemical evolution is characterized by a strong bias towards low-
and intermediate-mass stars, therefore its production of
$\alpha$-elements is strongly reduced.  The oxygen abundance smoothly
increases during the first $\sim$ 180 Myr (approximately up to the
break-out), then it stays almost constant.  As stressed in
Sect.~\ref{results_ICSV_chem}, this is mainly due to the dilution
effect of the clouds (as we have seen, the differential loss of metals
is very limited in this model).  The final abundance is 12 + log (O/H)
$\simeq$ 6.4, $\sim$ 0.3 dex less than model ICSV.  Although in this
model also the number of stars in the interval 4 -- 7 M$_\odot$
(i.e. the main producers of primary nitrogen) is reduced, the final
log (N/O) is much larger ($\sim$ 0.4 -- 0.5 dex) compared to model
ICSV.  Finally, due to the fact that both massive and low-mass stars
contribute to the final carbon abundance, the final log (C/O) does not
deviate much from the value found in model ICSV, being only $\sim$ 0.1
dex larger.

\subsection{Model NCSM}
\label{results_NCSM}

Model NCSM has an initial setup aimed at reproducing the gross
characteristics of NGC1569.  For this model we assume, according to
the work of Angeretti et al. (\cite{ang05}) three episodes of SF: a
most recent one occurred between 37 and 13 Myr ago, at a rate of 0.13
M$_\odot$ yr$^{-1}$; an intermediate episode, commencing 150 Myr ago
and finishing 40 Myr ago, at a rate of 0.04 M$_\odot$ yr$^{-1}$ and an
older episode of SF, ending 300 Myr ago (therefore implying 150 Myr of
inactivity between this episode and the intermediate one) and
commencing 600 Myr ago, at a rate of 0.05 M$_\odot$ yr$^{-1}$.  The
setup is the same as the model NGC -- 5 described in Paper II, namely
the total mass inside the galaxy is 1.8 $\times$ 10$^8$ M$_\odot$, but
we add continuously created clouds.  At variance with the models ICSV
and ICKV, the clouds are created every 5 Myr, therefore their masses
are between 2 and 6.5 $\times$ 10$^5$ M$_\odot$, not far from the
values actually observed in the complex of \hi clouds spiraling around
NGC1569 (M\"uhle et al. \cite{mue05}) and significantly larger than the
ones considered in the previous model with continuous creation of
clouds.  Moreover, due to the assumption made in Paper II about the
nucleosynthetic prescriptions, also in this case we consider yields
(from both massive and IMS) taken from Meynet \& Maeder (\cite{mm02}).

\subsubsection{Dynamical results}
\label{results_NCSM_dyna}

\begin{figure*}[ht]
 \begin{center}
 \includegraphics[width=\textwidth]{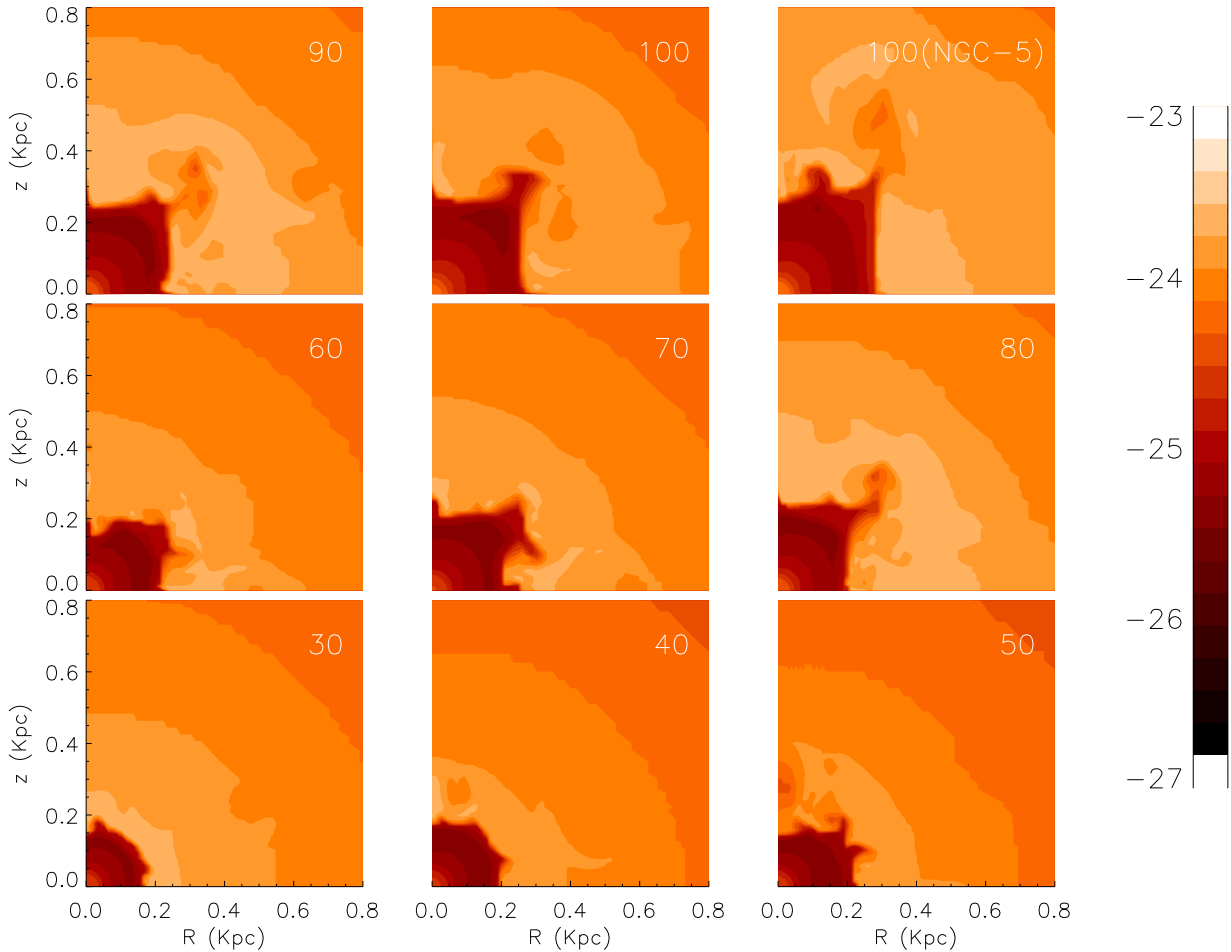}
\caption{ Same as Fig.~\ref{9s} but for model NCSM.  For reference, the 
density contours of model NGC -- 5 (presented in Paper II) are shown in 
the upper right panel.}
\label{ncsm} % for cross-references 
\end{center}
\end{figure*}

The dynamical evolution of this model is affected by the larger masses
of the clouds and by their lower creation rate.  It is therefore less
frequent the formation of clouds close in space and time, able
therefore to mutually interact.  The evolution of the model in the
first $\sim$ 100 Myr is shown in Fig.~\ref{ncsm}.  The impact of these
clouds on the development of large-scale outflows is stronger than in
the above-presented cases.  We can clearly see the collision of a
cloud with the expanding supershell in the lower right panel of this
figure.  These massive clouds lead also to a stronger evaporation
rate, therefore they affect significantly the energy budget of the
model.  The total thermal energy is $\sim$ 35 -- 40\% smaller than
the one attained by model NGC -- 5 (depending on the evolutionary
time).  Again, as we have noticed in Sect.~\ref{results_ICSV_dyna}, in
spite of the reduced superbubble power, the supershell-cloud
interaction can create holes in the supershell from which highly
pressurized gas can escape.  This is visible for instance in the
central right and in the upper central panel of Fig.~\ref{ncsm}.  At
$\sim$ 100 Myr the dimension of the superbubble is only slightly
smaller than the one attained at the same evolutionary time by model
NGC -- 5.  This demonstrates once again that the superbubble
luminosity is just one factor, in many cases not the leading one, in
determining the development and the shape of large-scale outflows.
Very important factors are also the density distribution and the
interaction with clouds.

\subsubsection{Chemical results}
\label{results_NCSM_chem}

Similarly to what is seen in Sect.~\ref{results_ICSV_chem}, the
dilution effect of the clouds overcomes the ``cap'' effect, due to the
delayed development of an outflow.  Therefore, model NCSM shows a
lower metallicity compared to model NGC -- 5.  In particular the
oxygen abundance, after an initial phase of $\sim$ 250 Myr in which it
grows constantly, remains in the range 12 + log(O/H) $\sim$ 7.6 --
7.7, therefore 0.3 -- 0.4 dex below the value attained by model NGC --
5.  Similarly to the diffuse model, the last intense burst of SF has
some effect on the chemical evolution of this model, resulting in an
increase of $\sim$ 0.1 -- 0.2 dex.  Also the log(N/O) is reduced by
$\sim$ 0.2 -- 0.3 dex compared to the corresponding diffuse model.
Since we had noticed in Paper II that the results model NGC -- 5
matched well the observed chemical composition of NGC1569 (taken from
Kobulnicky \& Skillman \cite{ks97}), we can point out that the
inclusion of a cloud complex {\it worsens} the agreement between model
results and observations.  In order to match the observed chemical
composition, one should therefore reduce the total initial mass, in
order to diminish the gas fraction and therefore increase the
metallicity.  Playing with this parameter is allowed by the present
uncertainties about the total mass of NGC1569 (Stil \& Israel
\cite{si02}; M\"uhle et al. \cite{mue03}), but, as already pointed out
in the Introduction, our main focus is not the quest for the best
setup able to reproduce the chemical composition of specific objects,
but the study of the effect of a cloud complex and the differences
with a model in which the gaseous distribution is smooth.

\section{Discussion and conclusions}
\label{discussion}

In this paper we have computed the chemical and dynamical evolution of
model galaxies, with structural parameters similar to IZw18 and
NGC1569, but in which a complex of clouds has been added, both
perturbing the initial gaseous distribution and creating clouds, at a
rate which equals the SF rate, and with infall velocity of 10 km
s$^{-1}$ along the polar direction.  The main focus of our work has
been the comparison of these models with those presented in previous
publications, in which similar setups but a smooth distribution of gas
was considered.

We have seen that the clouds are subject to a variety of disruptive
phenomena like evaporation (when embedded in a hot medium), formation
of shocks, development of thermal instabilities (in particular the
Kelvin-Helmholtz instability) and expansion due to the larger pressure
compared to the surrounding interstellar medium.  The average lifetime
of the clouds is therefore relatively short, depending on the cloud
size (which is not constant in our simulations) but being of the order
of a few tens of Myr.  In spite of their transient nature, the clouds
leave a significant imprint on the dynamical and chemical evolution of
dwarf galaxies.  The clouds, when they evaporate inside the
superbubble, produce mass loading, increase the mean density of the
cavity and, therefore, enhance the radiative losses (which are
proportional to the square of the density).  This results in a
significant decrease of the total thermal energy (of the order of
$\sim$ 20 -- 40\% compared to the diffuse models, depending on the
assumptions), therefore less energy to drive the development of a
large-scale outflow.  

On the other hand, the relative motion of supershell and clouds, in
particular when the clouds infall motion is considered, can structure,
pierce and create holes and fingers in the expanding supershell.
These holes destroy the spherical symmetry initially present and favor
the rushing out of the highly pressurized gas contained in the cavity.
Therefore, in spite of the reduced thermal energy budget, the creation
of large-scale outflows is not suppressed but, in most of the explored
cases, only slightly delayed.  Complex structures and fingers are
indeed relatively common features in galaxies showing large-scale
outflows like NGC1800 (Hunter \cite{hun96}), NGC4214 (MacKenty et
al. \cite{mack00}) or NGC1705 (Heckman et al. \cite{hek01}).  The
pressure inside the cavity is reduced compared to diffuse models,
therefore in any case the total amount of ejected pristine gas is very
small (smaller than in the models with smooth gas distribution) and,
when averaging the size of the supershell in any direction, it turns
out to be smaller than in diffuse models.  But the piercing of the
supershell can lead to an ejection efficiency of freshly produced
metals as high as the one attained by diffuse models.

%COMMENT: WHICH PRISTINE GAS? THE EVAPOPRATED CLOUD MATERIAL?

This has, of course, important consequences on the chemical evolution
of these objects.  Since the differential winds are not suppressed,
the diminished thermal energy of these models does not imply an increase
of metals inside the galactic regions.  On the other hand, the
dilution effect of clouds plays a dominant role in determining the
final metallicity of our model galaxies.  Since the clouds have
primordial chemical composition, their destruction and mixing with the
surrounding medium reduces the total chemical composition without
altering the abundance ratios.  This produces a final metallicity
$\sim$ 0.2 -- 0.4 dex smaller than the corresponding diffuse models.

We have examined the effect of a different choice of the IMF slope and
of the nucleosynthetic set of yields (in particular for what concerns
intermediate-mass stars).  Flatter-than-Salpeter IMF slopes lead to an
excessive production of energy, able to unbind most of the gas before
the end of the simulation.  On the other hand, in models with steeper
IMF the development of large-scale outflows is almost completely
suppressed.  Different sets of intermediate-mass stars yields affect
in particular the log(N/O) ratio.  Renzini \& Voli (\cite{rv81})
yields tend to overestimate the primary production of nitrogen.  When
compared to the results of models implementing van den Hoek \&
Groenewegen (\cite{vg97}) yields, the results differ by $\sim$ 0.3
dex.  Due to the assumption of a metallicity-dependent cooling
function, also the dynamics is affected by the choice of the
nucleosynthetic prescriptions.

Our main results can be briefly summarized as follows:

\begin{itemize}

\item the clouds suffer thermal instabilities, formation of shocks and
evaporation, therefore their lifetimes is limited to a few tens of
Myr.

\item In spite of that, they are able to increase the main density of the 
cavity, provoking a reduction of the total thermal energy by $\sim$ 20 -- 
40\% compared with a diffuse model.

\item The interaction clouds-supershell leads to strong structuring and 
piercing of the shell (in particular for models with continuous creation 
of infalling clouds), allowing the venting out of metals in spite of the 
reduced thermal energy.  The development of large-scale outflows is 
therefore generally delayed but the ejection efficiency of metals remains 
unchanged.

\item From a chemical point of view, the effect of the clouds is to 
significantly reduce the total metallicity of the galaxies, without 
altering the abundance ratios.

\end{itemize}

\begin{acknowledgements}
  
  We warmly thank the referee, Peter Berczik, for his suggestions that
  much improved the final version of this paper.  S.R. acknowledges
  generous financial support from the Alexander von Humboldt
  Foundation and Deutsche Forschungsgemeinschaft (DFG) under grant HE
  1487/28-1.

\end{acknowledgements}

\end{document}